# A Stable On-line Algorithm for Energy Efficient Multi-user Scheduling

Nitin Salodkar, Abhay Karandikar, *Member, IEEE* and V. S. Borkar, *Fellow, IEEE*

*Abstract*—In this paper, we consider the problem of energy efficient uplink scheduling with delay constraint for a multi-user wireless system. We address this problem within the framework of constrained Markov decision processes (CMDPs) wherein one seeks to minimize one cost (average power) subject to a hard constraint on another (average delay). We do not assume the arrival and channel statistics to be known. To handle state space explosion and informational constraints, we split the problem into individual CMDPs for the users, coupled through their Lagrange multipliers; and a user selection problem at the base station. To address the issue of unknown channel and arrival statistics, we propose a reinforcement learning algorithm. The users use this learning algorithm to determine the rate at which they wish to transmit in a slot and communicate this to the base station. The base station then schedules the user with the highest rate in a slot. We analyze convergence, stability and optimality properties of the algorithm. We also demonstrate the efficacy of the algorithm through simulations within IEEE 802.16 system.

## I. INTRODUCTION

Broadband wireless networks like IEEE 802.16 [1] and 3G cellular [2] are expected to provide Quality of Service (QoS) for emerging multimedia applications. One of the challenges in providing QoS is the time varying nature of the wireless channel due to multipath fading [3]. Moreover, for portable and hand-held devices, energy efficiency is also an important consideration.

For most wireless communication systems, the power required to transmit '*reliably*' for a given channel fading state is an increasing and strictly convex function of the transmission rate [3]. This suggests that energy efficiency can be achieved by transmitting the data at lower rates when the channel is bad, albeit at a cost of queuing delay, thus leading to a power-delay tradeoff. Furthermore in a multi-user wireless system, recent studies [4], [5] suggest that since the wireless channel fades independently across different users, this diversity can be exploited by *opportunistically* scheduling

Nitin Salodkar is currently with General Motors India Science Lab, ITPL, Bangalore India - 560066. Email: nitin.salodkar@gm.com Work done when Nitin was a doctoral candidate with Department of Computer Science and Engineering, IIT Bombay.

Abhay Karandikar is with Department of Electrical Engineering, IIT Bombay, Powai, Mumbai India - 400076. Email: karandi@ee.iitb.ac.in Work supported in part by Tata Teleservices-IIT Bombay Center for Excellence in Telecom.

V. S. Borkar is with the School of Technology and Computer Science, Tata Institute of Fundamental Research, Homi Bhabha Road, Mumbai, India - 400005. Email: borkar@tifr.res.in Work supported in part by J. C. Bose Fellowship.

the user with the best channel gain. This leads to significant performance improvement in terms of total system throughput. Such scheduling algorithms that exploit the characteristics of the physical channel to satisfy some network level QoS performance metrics are referred to as *cross layer* scheduling algorithms [6].

In this paper, we consider a single cell multi-user wireless uplink system. For such a system, we consider the problem of determining the user to be scheduled in each time slot along with its transmission rate. This scenario may correspond to scheduling users on the uplink in an IEEE 802.16 based system to satisfy delay constraint of each user.

### A. Related Work

*1) Energy Efficient Scheduling:* The problem of energy efficient scheduling with delay constraint for a single user wireless channel has been explored in the pioneering work of [7]. Subsequently, the model of Berry-Gallager [7] has been extended with many generalizations on arrival and channel state processes in [8], [9], [10], [11], [12], [13], [14]. In most of these papers, the scheduling policy has been formulated as a control policy within the Constrained Markov Decision Process (CMDP) framework. However, only structural results of the optimal policy are available under various assumptions. Moreover, these are applicable to only the single user scenario. There is very little work for extending the vast body of literature on single user delay constrained energy efficient scheduling to the multi-user scenario.

In [15], the author does extend the analysis for single user case to multi-user case, albeit with only two users. Beyond two users, the problem becomes too unwieldy to gain any useful insight. This is primarily due to the large state space. For the two user case, the author has given an elegant near optimal policy where each user's rate allocation is determined by the joint channel states across users and the user's own queue state. Thus each user's queue evolution process behaves as if it were controlled by a single user policy. However, computation of user's transmission power still takes into account the joint channel and queue state processes.

Recently, in a significant work [16], the author has extended the asymptotic analysis of Berry-Gallager [7] for exploiting the power-delay tradeoff to a multi-user system. The author, however, has considered the case of downlink scheduling, i.e, the base station scheduling users on the downlink. The objective is to minimize the total sum power subject to the users' queue stability constraints. Using the concept of Lyapunov Drift Steering, the author has given an algorithm that comes within a logarithmic factor of achieving the Berry-Gallager power-delay bound. While [16] is one of the first serious attempts at multi-user energy efficient scheduling with delay constraint, it deals with sum power minimization on the downlink. On the other hand, for the uplink, the problem is to minimize the average power of *each user* subject to *individual delay* constraint, subject to the additional constraints automatically imposed by the multi-user environment. This has not been addressed in the literature so far.

Even for the single user case [8], [9], [10], [11], [12], [13], [14], practical implementation of optimal policy is far from simple. This is because a knowledge of the probability distributions of the arrival and channel state processes is required for computing the optimal policy. This knowledge is usually not available in practice. We have addressed this limitation by formulating an on-line algorithm within stochastic approximation framework in [17].

*2) Other Multi-user Cross Layer Scheduling Schemes:* While there has been little work in the area of multi-user energy efficient delay constrained scheduling, there is an abundance of literature on cross layer scheduling algorithms with other objectives. See [18] for a succinct review. A scheduling policy is considered *stable* if the expected queue lengths are bounded under the policy. Many scheduling policies proposed in the literature have considered stability as a QoS criterion. In [19], the authors have shown that the throughput capacity region (as derived in [4]) is the same as the multi-access stability region (i.e., the set of all arrival vectors for which there exists some rate and power allocation policies that keep the system stable.). A scheduler is termed *throughput optimal* if it can maintain the stability of the system as long as the arrival rate is within the stability region. Throughput optimal scheduling policies have been explored in [4], [20]. Longest Connected Queue (LCQ) [21], Exponential (EXP) [22], Longest Weighted Queue Highest Possible Rate (LWQHPR) [23] and Modified Longest Weighted Delay First (M-LWDF) [24] are other well known throughput optimal scheduling policies.

While throughput optimal scheduling policies maintain the stability of the queueing system, they do not necessarily guarantee small queue length and consequently lower delay. Delay optimal scheduling deals with optimal rate and power allocation such that the average queue length and hence average delay are minimized for arrival rates within the stability region under average and peak power constraints. It has been shown that the Longest Queue Highest Possible Rate (LQHPR) policy [25] (besides being throughput optimal) is also delay optimal for any symmetric power control under symmetric fading provided that the packet arrival process is Poisson and the packet length is exponentially distributed.

Apart from throughput and delay optimal policies, opportunistic scheduling which maximizes sum throughput subject to various fairness constraints have been explored in [26], [27].

### B. Our Contributions

In this paper, we consider the problem of opportunistic scheduling for a multi-user uplink system with power cost and individual delay constraints. Considered as a centralized control problem of power minimization subject to constraints on average delays, this would be a special case of a *constrained Markov decision process* (CMDP). The traditional approaches for numerically determining the optimal policy in a CMDP framework are based on Linear Programming (LP) [28]. These, however, cannot be used for the problem posed in this paper because of the following reasons:

1) *Large state space:* In our model, the system state space is large even for moderate number of users and the state space size increases exponentially with the number of users. We illustrate this with a simple example. Consider a system with 4 users. Assume that each user has a buffer of size 50 packets (assuming equal sized packets). The channel condition of each user can be represented using 8 states, which is a practical assumption justified in [29]. For this scenario, the system state space contains $50^4 \times 8^4 = 2.56 \times 10^{10}$ states. The computational complexity for determining the optimal policy (possibly based on the CMDP approach) is proportional to the state space size [30], [31] and thus increases exponentially with the number of users.

2) *Unknown system model:* Computation of optimal scheduling policy using traditional schemes based on LP assumes a knowledge of the system model, i.e., a knowledge of probability distributions of the arrival and channel state processes, for modeling the transition probability mechanism of the underlying Markov chain. This knowledge is not easily available in practice, so the exact model is not known.

3) *Communication overheads:* In the multi-user framework considered here, there is also a cost on messages communicated between users and base station, as these consume some of the available rate. Thus any proposed scheme should be low on these overheads, which works against any centralized scheme based on full state information.

The issue of unknown system model can be resolved by using reinforcement learning (RL) algorithms [31] which 'learn' the optimal policy by performing approximate dynamic programming based on observed data. However, with such a large state space, the learning algorithms would take prohibitively large time to converge to the optimal scheduling policy. One therefore has to address the issue of the large state space first and then employ the reinforcement learning algorithms appropriately. This provides us the motivation for designing multi-user scheduling policy as a combination of single user policies that search over a relatively small state space. This is achieved by artificially splitting the problem into several single user problems for the individual users and the base station, which are coupled, but in a relatively simple and manageable manner. This reduces the complexity to linear in the number of users.

Thus in our approach, each user behaves as though it is facing a single user optimization problem and comes up with a desired rate. This is then communicated to the base station. The base station then schedules the user with the highest rate requirement in a slot. The intuition behind this is that this will favor the user with greatest need, be it because of a favorable channel or a high queue length. At the same time, when a user with lower rate requirement is not allocated the channel for a while, its queue length and therefore the rate requirement will go up and it will eventually be allocated the channel. The learning algorithm puts a penalty on violation of the delay constraint. This implicitly couples individual decisions, as the users are sharing a common channel.

The scheduling algorithms proposed in the literature like EXP [22], LQHPR [25], M-LWDF [24] etc., also require the queue length information for determining the scheduling decision. In the downlink scenario, this information is readily available to the scheduler residing at the base station. However, in the uplink scenario, this information needs to be communicated by the users to the scheduler. Communicating the queue length information poses a significant overhead. In our approach, each user needs to communicate only the desired rate. In a practical system, we may have few possible rates, say eight. This means that we may need only 3 bits of information to be conveyed.

Through our simulations in an IEEE 802.16 system, we demonstrate that the algorithm is indeed able to satisfy the delay constraints of the users. Moreover, we demonstrate that the power expenditure of a user is commensurate with its delay requirement, average arrival rate and average channel condition.

The contributions of this paper can be summarized as follows:

1) We propose a novel scheduling algorithm for minimizing the average power expended by each user subject to a constraint on individual user delay in a multi-user uplink wireless system. This algorithm does not require knowledge of the probability distributions of the channel states and the arrivals.
2) We analyze convergence, stability and optimality properties of the proposed scheme. In particular, we show certain desirable properties such as Pareto optimality and an interpretation as 'Markov equilibrium' of a stochastic game. We also argue incentive compatibility of the scheme.
3) We demonstrate applicability of the algorithm within IEEE 802.16 framework. Our simulation studies involving comparison with M-LWDF scheduler demonstrate that the proposed algorithm is power efficient. We also study performance of the scheme under different 'information accuracies', i.e., with different number of bits for conveying the desired rate.

The rest of the paper is organized as follows. In Section II, we present the system model. In Section III, we formulate the multi-user scheduling problem. In Section IV, we propose an on-line learning algorithm for the users and a user selection algorithm at the base station. We also discuss the implementation issues. In Section V we analyze properties of the algorithm such as convergence, queue stability and optimality.

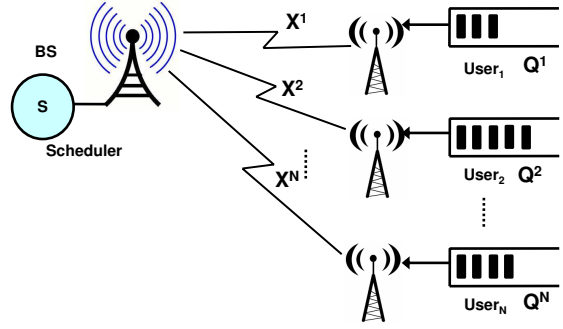

Fig. 1.   System Model

We present the simulation setup and results in Section VI. Finally, we conclude in Section VII.

## II. System Model

We consider uplink transmissions (as in Figure 1) in a Time Division Multiple Access (TDMA) system[1] with $N$ users, i.e., time is divided into slots of equal duration and only one user is allowed to transmit in a slot. We assume that the slot duration is normalized to unity. The base station is a centralized entity that schedules the users in every slot. We assume a fading wireless channel where the channel gain is assumed to remain constant for the duration of the slot and to change in an independent and identically distributed (i.i.d.) manner across slots. This model is termed as the *block fading* model [7]. We assume that the channel gains across users are also i.i.d. Under these assumptions, if a user $i$ transmits a signal $y_n^i$ in slot $n$, then the received signal $Z_n^i$ can be expressed as,

$$Z_n^i = H_n^i y_n^i + G_n, \qquad (1)$$

where $H_n^i$ denotes the complex channel gain due to fading and $G_n$ denotes the complex additive white Gaussian noise with zero mean and variance $N_0$. Let $X_n^i = |H_n^i|^2$ be the channel state for user $i$ in slot $n$. Usually, $H_n^i$ (and hence $X_n^i$) is a continuous random variable. However, in this paper we assume that $X_n^i$ takes only finite and discrete values from a set $\mathcal{X}$. This assumption has been justified in [7], [8]. In this paper, we assume that the distribution of $X_n^i$ is unknown.

We assume that the users' packets are of equal size, say, $\tau$ bits. Packets arrive into the user buffer of infinite capacity and are queued until they are transmitted. The packet arrival process for each user is assumed to be i.i.d. across slots. Let $A_n^i$ denote the number of packets arriving into the user $i$ buffer in slot $n$. We assume that the random variable $A_n^i$ takes values from a finite and discrete set $\mathcal{A} \triangleq \{0, \ldots, A\}$. Like $X_n^i$, we assume that the distribution of $A_n^i$ is unknown.

Let $Q_n^i \in \mathcal{Q} \triangleq \{0, 1, \ldots\}$ denote the queue length or buffer occupancy of user $i$ in slot $n$. Let $U_n^i$ denote the number of packets actually transmitted in slot $n$ (by user $i$). We assume that $U_n^i$ takes values from the set $\mathcal{U} \triangleq \{0, 1, \ldots\}$. Let $I_n^i$ be an indicator variable that is set to 1 if user $i$ is scheduled in slot $n$ and is set to 0 otherwise. Let $\mathbf{I}_n$ be the vector $[I_n^1, \ldots, I_n^N]$. Note that since only one user can transmit in a slot, only one element of $\mathbf{I}_n$ is equal to 1 and the rest are 0. Let $\mathcal{I}$ be the set of all possible $N$ dimensional vectors with one element equal to 1 and the rest being 0. Let $R_n^i \in \mathcal{U}$ denote the number of packets that user $i$ should transmit in a slot if it is scheduled. Then $U_n^i$ can be represented as $U_n^i = I_n^i R_n^i$. Moreover, since a user can at most transmit all the packets in its buffer in a slot, $R_n^i \leq Q_n^i$. Since we assume that the slot length is normalized to unity, $U_n^i$ also represents the rate at which user $i$ transmits in slot $n$. Let $\mathbf{U}_n$ be the vector $[U_n^1, \ldots, U_n^N]$, $\mathbf{U}_n \in \mathcal{U}^N$.

The queue evolution equation for user $i$ can be expressed as,

$$Q_{n+1}^i = Q_n^i - U_n^i + A_{n+1}^i. \quad (2)$$

For most communication systems, the power required for reliable transmission at a rate $R_n^i = u$ packets/sec when $X_n^i = x$, denoted as $P(x, u)$, is an increasing and strictly convex function of $u$. Let $\hat{P}$ denote the maximum power with which a user can transmit in any slot. Let $\hat{R}^i(x^i)$ be the maximum number of packets which user $i$ can transmit in a slot when the channel state is $x^i$ when transmitting with power $\hat{P}$. Then the set of feasible actions for user $i$ in slot $n$, $\mathcal{F}_n^i \triangleq \{0, \ldots, \min(\hat{R}^i(x_n^i), Q_n^i)\}$.

We assume that the users specify their QoS requirements in terms of the average packet delay requirements. These delay requirements of the users are known a-priori to the scheduler. By Little's law [32], the average delay $\bar{D}$ is related to the average queue length $\bar{Q}$ as,

$$\bar{Q} = \bar{a}\bar{D}, \quad (3)$$

where $\bar{a}$ is the average arrival rate. In the rest of the paper, we treat average delay as synonymous with average queue length and ignore the proportionality constant $\bar{a}$.

## III. Problem Formulation

In this section, we formulate the multi-user scheduling problem.

### A. Formulation as a Constrained Optimization Problem

The problem considered here is to design a scheme for scheduling a user in each time slot and also the rate (i.e., number of packets to be transmitted) that minimizes the average power expenditure of each user subject to the satisfaction of individual delay constraint. The average power consumed by a user $i$ can be expressed as:

$$\bar{P}^i = \limsup_{M \to \infty} \frac{1}{M} \mathbf{E} \sum_{n=1}^{M} P(X_n^i, I_n^i R_n^i). \quad (4)$$

The average queue length of user $i$ can be expressed as:

$$\bar{Q}^i = \limsup_{M \to \infty} \frac{1}{M} \mathbf{E} \sum_{n=1}^{M} Q_n^i. \quad (5)$$

Each user $i$ wants its average queue length to remain below a certain value, say, $\bar{\delta}^i$. Our objective is to design a scheduling algorithm that minimizes $\bar{P}^i$ for each user $i$ subject to a constraint on $\bar{Q}^i$. Thus the scheduler objectives can be stated as,

$$\text{Minimize } \bar{P}^i \text{ subject to } \bar{Q}^i \leq \bar{\delta}^i, \ i = 1, \ldots, N. \quad (6)$$

*Remark 1:* Note that there are actually $N$ problems in (6). However, these problems are not independent. This is because in a TDMA system, only one user can be scheduled in a slot. Consequently, the scheduling decision in a slot impacts the buffer occupancy of all the users in future slots.

### B. Notion of an Optimal Solution

The problem in (6) is an optimization problem with $N$ objectives and $N$ constraints. There can be multiple average power vectors that can be considered as optimal. Let $[\bar{P}_\Phi^1, \ldots, \bar{P}_\Phi^N]$ denote the average power expended under the scheduling policy $\Phi$. We say that the scheduling policy $\Phi$ is Pareto optimal if and only if there exists no policy $\vartheta$ with the corresponding power expenditure vector $[\bar{P}_\vartheta^1, \ldots, \bar{P}_\vartheta^N]$ having the following properties,

$$\forall i \in \{1, \ldots, N\} \bar{P}_\vartheta^i \leq \bar{P}_\Phi^i \ \wedge \ \exists i \in \{1, \ldots, N\} | \bar{P}_\vartheta^i < \bar{P}_\Phi^i. \quad (7)$$

We seek a Pareto optimal solution in this paper.

In the next section, we present a solution strategy wherein we split the problem into $N + 1$ parts: $N$ user problems and one base station problem.

## IV. Solution Strategy: Decomposition into User and Base Station Problems

In this section, we propose the following solution strategy: we view the problem as $N$ (dependent) user problems and one base station problem. The user $i$ problem is to determine a rate at which it desires to transmit in a slot so as to solve the problem specified in (6). Since the channel and arrival statistics are not known, in order to address this problem, the users resort to a 'learning' approach discussed below. The users' desired rates are then conveyed to the base station. The base station problem is to select a user in each slot such that the average power expenditure vector is a point on the Pareto frontier, i.e., it is Pareto optimal. Next, we present the on-line learning algorithm at users.

### A. Learning Algorithm for the Users

We consider a modified version of the on-line algorithm proposed in [17] that determines the transmission rate in every slot for each user. Once the on-line algorithm has determined the rate $R_n^i \in \mathcal{F}_n^i$, the transmitter at a particular user executes this action if the channel is allocated to it, otherwise it is unable to proceed with the transmission. If the transmitter is not able to proceed with the transmission, the packets remain in the queue. Under the given model, the queue evolution equation can be expressed as,

$$Q_{n+1}^i = Q_n^i + A_{n+1}^i - I_n^i R_n^i, \quad (8)$$

where $I_n^i = 1$ if the transmitter $i$ actually transmits the packets (i.e., if user $i$ is scheduled in a slot $n$), else, $I_n^i = 0$. Let $\mathbf{Q}_n$ and $\mathbf{X}_n$ denote the vectors $[Q_n^1, \ldots, Q_n^N]$ and $[X_n^1, \ldots, X_n^N]$ respectively. The state of the system $\mathbf{S}_n$ at time $n$ can be described by $\mathbf{S}_n = (\mathbf{Q}_n, \mathbf{X}_n)$ comprising of the queue lengths and the channel states. The control variables are $\mathbf{I}_n$, $\mathbf{R}_n = [R_n^1, \ldots, R_n^N]$, of which the transmitters independently choose the corresponding components of $\mathbf{R}_n$ and the base station, which is doing the channel allocation, chooses $\mathbf{I}_n$ subject to the constraints that $I_n^i \in \{0, 1\}$ and $\sum_i I_n^i = 1$.

Our learning policy, however, mandates that once transmitter $i$ has determined the rate $R_n^i$, it updates its power cost irrespective of whether packets have been transmitted or not. That is, it operates as though its cost is

$$\bar{P}_e^i = \limsup_{M \to \infty} \frac{1}{M} \mathbf{E} \sum_{n=1}^{M} P(X_n^i, R_n^i). \qquad (9)$$

Thus the problem that user $i$ addresses can be specified as:

$$\text{Minimize } \bar{P}_e^i \text{ subject to } \bar{Q}^i \le \bar{\delta}^i, \ \ i = 1, \ldots, N. \qquad (10)$$

This problem is a CMDP with average cost criterion. The objective is to determine an optimal policy $\mu^{i,*}$ such that the power cost (9) is minimized while satisfying the delay constraint. Note that the policy considered here minimizes the single user cost exactly as in the single user case of [17] as specified in (9), not the actual power cost (4). See the Remark 2 at the end of this sub-section and the subsequent sections for further discussion on this aspect.

*1) The Primal Dual Approach:* The constrained problem in (10) can be converted into an unconstrained problem using the Lagrangian approach [28]. We focus on the $i$th user. Let $\lambda^i \ge 0$ be a real number termed as the Lagrange Multiplier (LM). Let $c : \mathcal{R}^+ \times \mathcal{Q} \times \mathcal{X} \times \mathcal{U} \to \mathcal{R}$ be defined as the following,

$$c(\lambda^i, Q_n^i, X_n^i, R_n^i) \triangleq P(X_n^i, R_n^i) + \lambda(Q_n^i - \bar{\delta}^i), \qquad (11)$$

where $R_n^i$ is determined using the rate allocation policy $\mu^i : \mathcal{Q} \times \mathcal{X} \to \mathcal{U}$. The unconstrained problem is to minimize:

$$L(\mu^i, \lambda^i) = \limsup_{M \to \infty} \frac{1}{M} \mathbf{E} \sum_{n=1}^{M} c(\lambda^i, Q_n^i, X_n^i, \mu^i(Q_n^i, X_n^i)). \qquad (12)$$

$L(\cdot, \cdot)$ is called the Lagrangian. Our objective is to determine the optimal rate allocation policy $\mu^{i,*}$ and optimal LM $\lambda^{i,*}$ such that the following saddle point optimality condition [33] is satisfied:

$$L(\mu^{i,*}, \lambda^i) \le L(\mu^{i,*}, \lambda^{i,*}) \le L(\mu^i, \lambda^{i,*}). \qquad (13)$$

For a fixed LM $\lambda^i$, the problem is an unconstrained Markov Decision Problem (MDP) with the average cost criterion. Let $p((q, x), r, (q', x'))$ denote the probability of reaching a state $(q', x')$ upon taking an action $r$ in state $(q, x)$. Let $V^{i,\mu}(\cdot, \cdot)$ denote the value function under policy $\mu$, i.e., $V^{i,\mu}(q, x)$ denotes the expected cost for a state $(q, x)$ under a policy $\mu$. Let $V^i(\cdot, \cdot)$ denote optimal value function. It can be expressed as:

$$V^i(q, x) = \min_{\mu} V^{i,\mu}(q, x). \qquad (14)$$

The following dynamic programming equation [30] gives a necessary condition for the optimality of a solution:

$$V^i(q, x) = \min_{r \in \mathcal{F}} \Big[ c(\lambda^i, q, x, r) - \beta + \sum_{q', x'} p((q, x), r, (q', x')) V^i(q', x') \Big], \qquad (15)$$

where $a' \in \mathcal{A}$, $x' \in \mathcal{X}$, $V^i(\cdot, \cdot)$ is the value function, $\beta \in \mathcal{R}$ is the unique optimal power expenditure and $(q^0, x^0) \in \mathcal{Q} \times \mathcal{X}$ is any pre-designated state. If we impose $V^i(q^0, x^0) = 0$, then $V^i(\cdot, \cdot)$ is unique [30]. The Relative Value Iteration Algorithm (RVIA) [30] is a known approach for determining the optimal value function. It can be expressed as:

$$V_{n+1}^i(q, x) = \min_{r \in \mathcal{F}^i} \Big[ c(\lambda^i, q, x, r) - V_n^i(q^0, x^0) + \sum_{q', x'} p((q, x), r, (q', x')) V_n^i(q', x'). \qquad (16)$$

Note, however, that RVIA (16) requires the knowledge of $p(\cdot, \cdot, \cdot)$. This depends on the probability distributions of the arrivals and channel states; which are not known. Moreover, determining the optimal value function as defined in (15) is not sufficient because the unconstrained solution for a particular $\lambda^i$ does not ensure that the constraints would be satisfied. To ensure constraint satisfaction, the optimal LM needs to be determined.

*2) The Post-decision State Formulation:* To address the difficulty posed by unknown $p(\cdot, \cdot, \cdot)$, we introduce a stochastic approximation version of the RVIA. The RVIA above, however, is not suited for this because the conditional expectation w.r.t. the transition probability is inside the (nonlinear) minimization operator. This prompts us to replace the state variables $(Q_n^i, X_n^i)$ by 'post-decision state' variables[2] $(\tilde{Q}_n^i, \tilde{X}_n^i)$ defined by $\tilde{Q}_n^i = Q_n^i - U_n^i, \tilde{X}_n^i = X_n^i$. Let $\zeta^i(\cdot)$ denote the unknown law for the arrivals and $\kappa^i(\cdot|\cdot)$ the transition probability kernel for the channel states. The dynamic programming equation and the RVIA for post-decision states become

$$\tilde{V}^i(\tilde{q}, \tilde{x}) = \sum_{a', x'} \zeta^i(a') \kappa(x'|x) \times \min_{r \in \mathcal{F}^i} \Big[ c(\lambda^i, \tilde{q} + a', x', r) - \beta + \tilde{V}^i(\tilde{q} + a' - r, x') \Big], \qquad (17)$$

and

$$\tilde{V}_{n+1}^i(\tilde{q}, \tilde{x}) = \sum_{a', x'} \zeta^i(a') \kappa(x'|x) \times \min_{r \in \mathcal{F}^i} \Big[ c(\lambda^i, \tilde{q} + a', x', r) - \tilde{V}_n^i(\tilde{q}^0, \tilde{x}^0) + \tilde{V}_n^i(\tilde{q} + a' - r, x') \Big], \qquad (18)$$

where $(\tilde{q}^0, \tilde{x}^0)$ is any pre-designated reference post-decision state. On the right hand side in (18), the value function corresponding to this state is subtracted in order to keep the iterates bounded. In the next section, we present an on-line algorithm based on a reformulation of this RVIA that does not require the knowledge of $p(\cdot, \cdot, \cdot)$. The approach is similar to that in [17]. We also augment the reformulated RVIA with an LM iteration to ensure constraint satisfaction.

---

[2]This is a special construct that is possible when the controlled transition naturally splits into the control action followed by the action of noise. It has the advantage that the corresponding dynamic programming equation has the conditional expectation operation outside the minimization operation. This facilitates a stochastic approximation version of it, which is the learning algorithm. See [34] for an extensive account of the post-decision state formalism.

*3) The On-line Rate Allocation Algorithm:* In this section, we present the on-line algorithm which we call the *rate allocation algorithm*. This algorithm determines the rate with which the user should transmit in a slot and updates the value function and LM.

Let $\{f_n\}$ and $\{e_n\}$ be two sequences that have the following properties,

$$\sum_n (f_n)^2, \quad \sum_n (e_n)^2 < \infty, \tag{19}$$

$$\sum_n f_n = \infty, \quad \sum_n e_n = \infty, \tag{20}$$

$$\lim_{n \to \infty} \frac{e_n}{f_n} \to 0. \tag{21}$$

The significance of these properties has been explained below. Let the user $i$'s state at the beginning of slot $n$ be $S_n^i = (Q_n^i, X_n^i) = (q, x)$. Suppose that $u$ packets are transmitted in slot $n$. The following primal-dual algorithm can be used to compute the rate $R_{n+1}^i = r_{n+1}^i$ at which the transmitter should transmit in slot $n + 1$,

$$
\begin{aligned}
r_{n+1}^i &= \arg \min_{v \in \mathcal{F}_{n+1}^i} \Big\{ (1 - f_n)\tilde{V}_n^i(\tilde{q}, \tilde{x}) + f_n \times \\
&\quad \Big\{ c(\lambda_n^i, \tilde{q} + A_{n+1}^i, X_{n+1}^i, v) \\
&\quad + \tilde{V}_n^i(\tilde{q} + A_{n+1}^i - v, X_{n+1}^i) \\
&\quad - \tilde{V}_n^i(\tilde{q}^0, \tilde{x}^0) \Big\} \Big\},
\end{aligned}
\tag{22}
$$

$$
\begin{aligned}
\tilde{V}_{n+1}^i(\tilde{q}, \tilde{x}) &= (1 - f_n)\tilde{V}_n^i(\tilde{q}, \tilde{x}) + f_n \times \\
&\quad \Big\{ c(\lambda_n^i, \tilde{q} + A_{n+1}^i, X_{n+1}^i, r_{n+1}^i) \\
&\quad + \tilde{V}_n^i(\tilde{q} + A_{n+1}^i - r_{n+1}^i, X_{n+1}^i) \\
&\quad - \tilde{V}_n^i(\tilde{q}^0, \tilde{x}^0) \Big\},
\end{aligned}
\tag{23}
$$

$$\lambda_{n+1}^i = \Gamma[\lambda_n^i + e_n \left( Q_n^i - \bar{\delta}^i \right)], \tag{24}$$

where $\Gamma[\cdot]$ in (24) is a projection operator that projects the $\lambda^i$ iterates in the interval $[0, \mathcal{K}]$ for $\mathcal{K} >> 1$. This is to ensure boundedness of these iterates. Components other than the $(\tilde{q}, \tilde{x})$th in (23) are left unchanged. These equations are explained below:

1) (22), (23) and (24) constitute the rate allocation algorithm or the 'user algorithm'. It consists of two phases: *rate determination phase* and *update phase*. (22) constitutes the rate determination phase of the algorithm, i.e., it is used to determine the rate at which a user should transmit in a slot. (23) is a primal iteration to determine the optimal value function for post-decision states and thereby the optimal policy, while (24) is a coupled dual iteration for determining the optimal LM. They constitute the update phase of the algorithm.

2) The sequences $f_n$ and $e_n$ have properties specified in (19), (20) and (21). The properties of the update sequences in (19) ensure that the sequences $\{f_n\}$ and $\{e_n\}$ converge to zero sufficiently fast to eliminate the noise effects when the iterates are close to their optimal values $\tilde{V}^i(\cdot, \cdot)$ and $\lambda^{i,*}$, while those in (20) ensure that they do not approach zero too rapidly to avoid convergence of the algorithm to non-optimal values. Furthermore, (21) ensures that the update rates of primal, i.e., the value function iterations and the dual, i.e., the LM iterations are different. Since $e_n$ approaches 0 much faster than $f_n$, the update rate of the value function iterations is much higher than that of the LM iterations. Using a two time scale analysis, we show in Section V-A that even though both the primal and dual variables are updated simultaneously, both converge to their optimal values.

*Remark 2:* Note that the r.h.s. of (23) is nothing but the actual minimum of the expression being minimized on the r.h.s. of (22). That is, the relative value iteration is being performed for the single user cost exactly as in the single user case of [17] specified by (9), not for the power cost (4). As we argue later, it will in fact converge to the single user optimum rate required *for the current quasi-static value of the Lagrange multiplier*. This is the rate the user will *request*, it is not necessarily the rate actually *allocated* to it. The difference with the single user case of [17] comes from the fact that these relative value iterations for users are coupled through the Lagrange multiplier updates which are indeed affected by the actual transmission scheme. As we see later, the convergence of (24) perforce implies that the constraints are met. Given the fact that not the full requested transmission rate is actually granted, that leads to the conclusion that a higher rate will be requested than in the single user case in [17], which in turn means larger limiting values for the $\lambda$'s compared to those in [17]. We call this the *multi-user penalty* that a user pays in the multi-user environment. Further discussion on this can be found in Section V-D1.

A convergence analysis of the user algorithm in (22), (23) and (24) is presented in Section V-A. Each user $i$ determines the rate $R_{n+1}^i = r_{n+1}^i$ at which it would transmit in slot $n + 1$ if it is scheduled, specified by the algorithm described above, and communicates this rate to the base station. The base station employs the user selection algorithm to schedule a user. The users update their value functions in each slot assuming successful transmission regardless of whether they are scheduled by the base station or not. Note, however, that for the user who is actually scheduled, a corresponding queue transition occurs because of the transmitted packets. This influences the queue length and consequently the LM update.

### B. User Selection Algorithm

The user selection algorithm schedules the user with the largest $R_n^i$. If more than one user has the largest rate then a user is selected at random from among them with uniform probability. The intuition behind selecting the user with the largest rate is the following. The rate allocation algorithm of a user $i$ would direct it to transmit at a high rate $R_n^i$ under two circumstances: either the channel condition for that user is very good, in which case transmission at high rate saves power,

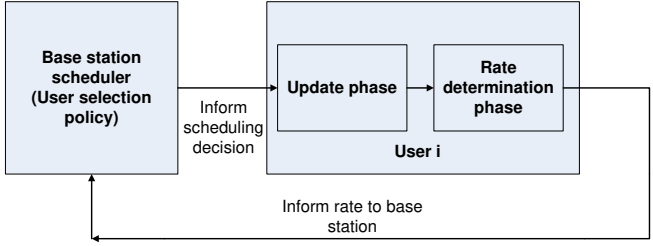

Fig. 2. Solution schematic

---

**1:** Initialize the value function matrix $\tilde{V}^i(q,x) \leftarrow 0 \quad \forall q \in \mathcal{Q}, x \in \mathcal{X}$
**2:** Initialize LM $\lambda^i \leftarrow 0$
**3:** Initialize slot counter $n \leftarrow 1$
**4:** Initialize queue length $q^i \leftarrow 0$
**5:** Initialize channel states $x^i \leftarrow 0, x^{i'} \leftarrow 0$
**6:** Reference state $= (0, x^1)$, where $x^1 \in \mathcal{X}$
**7: while** TRUE **do**
**8:**    **while** Base station has not informed the channel state $X^i_{n+1} = x^{i'}$ **do**
**9:**       wait
**10:**    **end while**
**11:**    Determine the number of arrivals $A^i_{n+1} = a^i$ in the current slot
**12:**    Determine the queue length in the current slot $Q^i_n = q^i$
**13:**    Use the rate determination phase of the rate allocation algorithm, i.e., (22) to determine the rate $r^i$, for transmission
**14:**    Determine the power $P(x^{i'}, r^i)$ required to transmit $r^i$ packets
**15:**    Inform the base station of the rate $r^i$
**16:**    **while** Base station has not scheduled a user **do**
**17:**       wait
**18:**    **end while**
**19:**    Update the component $(q^i, x^i)$ of the value function $\tilde{V}^i$ using (23). Rest of the components remain unchanged
**20:**    Update the LM $\lambda^i$ using (24) $(Q^i_n = q^i)$
**21:**    $q^i \leftarrow q^i + a^i - u^i$
**22:**    $x^i \leftarrow x^{i'}$
**23:**    $n \leftarrow n + 1$
**24: end while**

**Algorithm 1:** The Rate Allocation Algorithm at the User $i$ Device

---

**1: while** TRUE **do**
**2:**    **for** $i \in 1, \ldots, N$ **do**
**3:**       Estimate the channel state $X^i_{n+1} = x^{i'}$ in the current slot for user $i$
**4:**       Inform $x^{i'}$ to user $i$
**5:**    **end for**
**6:**    **while** Rate of each user is not known **do**
**7:**       wait
**8:**    **end while**
**9:**    Determine the user $k$ who has the highest rate
**10:**    Schedule user $k$ in the current slot
**11: end while**

**Algorithm 2:** The User Selection Algorithm at the Base Station

---

or the queue length of that user is large. Thus selecting a user with a high rate results in either power savings or reducing the queue length of that user and thereby ensuring that the delay constraint of that user is satisfied. An analysis of the user selection algorithm is presented in Section V-C1.

### C. Algorithm Details and Implementation

The rate allocation algorithm is implemented on the user devices while the user selection algorithm is implemented at the base station, as illustrated in Figure 2. From (22), note that the rate determination phase requires $X^i_n$, i.e., the knowledge of the channel state at the base station. The communication overhead incurred by the base station in informing a user about the channel state perceived by it depends on the number of states used to represent the channel. We represent the channel using 8 states. Thus the base station needs 3 bits per slot in order to inform a user about the channel state perceived by it. The users inform the base station about the computed rates. We allocate 3 bits for conveying this information, i.e., the system can employ 8 rates. The user selection algorithm then determines the user to be scheduled and all users are informed about this decision. The rate allocation algorithm at each user then enters the update phase where the value function and the LM for each user are appropriately updated using (23) and (24). The algorithm thus continues in each slot $n$. The rate allocation algorithm that is executed at each user device is illustrated in Algorithm 1 where, steps $8 - 13$ represent the rate determination phase, while steps $19 - 24$ represent the update phase. The user selection algorithm executed at the base station is detailed in Algorithm 2.

### D. Discussion

1) *Computational complexity:* The computational complexity of the rate allocation algorithm executed at a user device is independent of the number of users in the system. This is because the rate allocation algorithm for any user $i$ is dependent on the user $i$ state $S^i$ only and is independent of the states of the other users. The user selection algorithm has to determine the maximum of $N$ numbers and hence is linear in $N$. Thus the computational complexity of the user selection algorithm grows only linearly with the number of users.

2) *An auction interpretation:* The above scheme can be interpreted as an auction, where the user selection algorithm auctions each time slot. The users bid in the form of their transmission rates to the user selection algorithm, which allocates the time slot to the user bidding the highest rate. The rate bid by a user is dependent on its channel state and queue length. If the channel state is good and/or the queue length is large, the user bids a high rate. This is because transmitting at a high rate when the channel state is good saves power, while doing so when the queue length is large aids in satisfying the delays. Note that the users do not bid unnecessarily high rates because that might result in higher power consumption. For a user, not winning an auction in a certain slot implies that other users either have better channel conditions or relatively higher queue length or both. If a user does not win the auction for a certain number of slots successively, its queue length grows thus forcing it to bid a higher rate.

## V. Analysis of the Multi-user Scheme

In this section, we present an analysis of the multi-user algorithm presented in the paper. Specifically, we comment on queue stability, convergence and optimality of the algorithm. We begin with certain conditions that are necessary for the convergence of the algorithm.

Let $\varsigma(s^i, n)$ be the number of times that user $i$ state $s^i$ is visited up to time $n$. The state process should satisfy the following property:

$$\liminf_{n \to \infty} \frac{\varsigma(s^i, n)}{n} > 0, \ \ \forall s^i, i. \tag{25}$$

(25), which is akin to positive recurrence for uncontrolled Markov chains, is a stability condition and is essential for proving the convergence of our algorithm. We shall prove that the queues are indeed stable, implying that (25) is indeed satisfied, *if we assume* that the users have already learnt their policies, i.e., the user learning component of the overall scheme has converged. On the other hand, this convergence in turn requires a priori proof of stability. This circular situation is very common in adaptive and learning control (see, e.g., [35]) and a procedure that assures its resolution is yet unknown. There are however certain heuristic approaches that work well in practice. In the present case, for example, one can impose an initial 'pure learning' phase wherein one employs a known stable strategy for a finite duration instead of the 'self-tuning' policy which bases its decision on the current guess for the value function, and then switches to the latter. This will ensure that the latter, when finally inducted, gives good guesses for the value functions so that the corresponding guesses for actions are close to optimal, in particular stable. Another possible solution is to use a known stabilizing policy when the state process blows up, i.e., crosses a very large threshold, and use the policy proposed herein otherwise. Our simulations, however, use the original scheme and the simulation results are quite promising.

### A. Convergence Analysis

Note that each user in the system has coupled iterations comprising of the value function and LM. We need to prove that these coupled iterations converge to an equilibrium set for each user. The main steps in the analysis are as follows:

1) We make an assumption on the stability of the queues under the closed loop scheme, as noted above.

2) We then analyze the convergence of the value function for each user for an almost constant value of the LM. Since the value function of each user is updated in each slot regardless of whether the user is scheduled in that slot or not; the value functions are decoupled across users. This is in the spirit of the decoupling of static formulations of network flow problems via the Lagrange multiplier as in [36], except that here it is *mandated* by our algorithm. The decoupling is facilitated by the fact that the users compute their value function as though the cost is (9) and not (4). The latter would have introduced a more direct dependence on other users, over and above that through the LM.

3) Finally, we prove that the LMs and the coupled iterates of the users also converge. Convergence of the LMs implies that the delay constraints are satisfied and vice versa. This implies that if there is sufficient capacity, the multi-user scheduling satisfies the delay constraints of all the users.

We now prove that the value functions converge.

*Theorem 1:* For the rate determination algorithm (22), (23) and (24), the iterates $(\bar{V}_n^i, \lambda_n^i)$ converge to the optimal values, i.e., $(\bar{V}_n^i, \lambda_n^i) \to (\bar{V}^i, \lambda^{i,*})$. Moreover, convergence to equilibrium implies that the delay constraints of the users are satisfied.

*Proof:* The proof for individual user's algorithm is similar to that of the single user algorithm in [17]. We sketch it in outline below, referring to the relevant portions of [37] for details. The arguments are based on the well known 'o.d.e.' approach for the analysis of stochastic approximation algorithms, wherein one looks at the algorithm as a noisy discretization of a limiting ordinary differential equation (o.d.e.) which can be written by inspection. One treats the 'learning parameters' or stepsizes as discrete time steps and compares the linearly interpolated iterates with the o.d.e. trajectory from some time on. The assumptions such as (19) and (20) on the stepsizes ensure (under suitable hypotheses) that errors due to both discretization and noise are asymptotically negligible and therefore the iterates a.s. track the asymptotic behavior of the o.d.e. See Chapter 2 of [37] for the general idea of proof. We spell out some details below that are specific to this paper.

1) Our requirement that $\frac{e_n}{f_n} \to 0$ induces two time scales, a fast one for (23) and a slow one for (24). Let $h^i(\bar{V}^i) = [h_{q,x}^i(\bar{V}^i)]$ be given by:

$$h_{q,x}^i(\bar{V}^i) = \sum_{a,x'} \zeta^i(a)\kappa^i(x'|x) \times \min_u [c(\lambda^i, q + a, x, u) + \bar{V}^i(q + a - u, x') - \bar{V}^i(q^0, x^0)],$$

where $(q^0, x^0)$ is any pre-designated state. Using the two time scale analysis in [37], Section 6.1, we first analyze (23) by freezing $\lambda_n^i \approx$ a constant $\lambda^i$ and considering the limiting o.d.e. for (23) given by

$$\dot{\bar{V}}^i(t) = \Lambda^i(t)(h^i(\bar{V}^i(t)) - \bar{V}^i(t)), \tag{26}$$

where $\Lambda^i(t)$ is a diagonal matrix with nonnegative elements summing to 1 on the diagonal. The diagonal elements of $\Lambda^i(t)$ reflect the relative frequency with which the system states $(q, x)$ are visited, and therefore the corresponding components of the iteration are updated. Barring the diagonal matrix $\Lambda^i(t)$, this is just what one expects. The occurrence of $\Lambda^i(t)$ is due to the asynchronous iteration – we update only one component of $\tilde{V}_n^i$ at a time. See [37], Section 7.2, for a description of how this comes about.

2) Suppose that the queues remain stable. Coupled with the recurrence of the Markov chain describing the channel state, this implies that the empirical frequency of any possible post-decision state value remains bounded away from zero with probability one. This implies that the relative frequencies with which the states $(q, x)$ are visited are of the same order of magnitude. Hence, the diagonal elements of $\Lambda^i(t)$ remain uniformly bounded away from zero (see [37], p. 87, for a more formal argument). One also needs that a particular user is scheduled frequently enough by the base station. Our stability assumption will be seen to ensure this automatically: a user not selected for long builds up a large queue and hence requests higher rate, which favors its selection thereafter. Were this not so, that queue would have become unstable. In fact this can still happen if the maximum permissible rate for one user is way less than that for others, in which case it will be starved of transmission opportunities most of the time and go unstable. We assume that the maximum allowable rates are sufficiently large for the arrival rates under consideration and comparable (among users) so that this does not happen, in fact we later impose a condition ((30) below) stronger than this.

3) We now prove that the value function converges to its optimal value $\tilde{V}^i$, i.e., $V_n^i \to \tilde{V}^i$.

*Lemma 1:* If the diagonal elements of $\Lambda^i(t)$ remain uniformly bounded away from zero, $V_n^i \to \tilde{V}^i$.

*Proof:* We first need to prove that the iterates remain bounded a.s. We adapt the arguments of Section 3.2, [37]. The situation is slightly different here, viz., we have a time-dependent matrix $\Lambda^i(t)$ on the r.h.s. of (26) whereas Section 3.2 of [37] analyzes a time-homogeneous case (compare (26) in the present paper and (3.2.1) in [37]). Nevertheless, it does not affect the argument as long as the diagonal elements of $\Lambda^i(t)$ are uniformly bounded away from zero. As in Section 3.2 of [37], (especially the development after (3.2.2)), consider a 'scaled limit' of the o.d.e. (26) wherein in the r.h.s. of (26) the function $h^i$ is replaced by $h_\infty^i$ defined by $h_\infty^i(x) \triangleq \lim_{a \uparrow \infty} \frac{h^i(ax)}{a}$. This corresponds to (26) again, but with $c(\cdot) \equiv 0$, i.e., the immediate cost function is set to zero. It can be shown as in [17], Lemma 1, that this scaled o.d.e. has the origin as the globally asymptotically stable equilibrium. This ensures a.s. boundedness of the iterates by the results of Section 3.2 of [37]. The intuition is as follows: if the iterates become unbounded along a subsequence, a scaled version thereof, suitably interpolated, begins to approximate the limiting o.d.e. above and therefore has to return towards the origin by the asymptotic stability of the limiting o.d.e. Since these differ from the original iterates only by a scale factor, one can argue that the original iterates themselves start moving towards a bounded set and therefore cannot blow up. Section 3.2 of [37] makes this intuition precise.

One can then argue as in [17] (development just before Lemma 1) to conclude that $\tilde{V}^i(t)$ converges to the solution $\tilde{V}^i$ corresponding to $\tilde{V}^i(q^0, x^0) = \beta$, where $\beta$ is the optimal average cost per stage. As in Section 2.1, [37], it then follows that $V_n^i \to \tilde{V}^i$ a.s. ∎

4) Note that the above analysis treats $\lambda_n^i \approx$ a constant, so what we have really proved (cf. Section 6.1, [37]) is that $\{\tilde{V}_n^i\}$ closely tracks $\{\tilde{V}^{i,\lambda_n^i}\}$, where $\tilde{V}^{i,\lambda^i}$ is $\tilde{V}^i$ with its $\lambda$-dependence made explicit. To be precise, $\tilde{V}_n^i - \tilde{V}^{i,\lambda_n} \to 0$ a.s. The following lemma then proves that the LM converges to its optimal value $\lambda^{i,*}$, and hence the pair $(\tilde{V}_n^i, \lambda_n^i)$ converges to $(\tilde{V}^i, \lambda^{i,*})$.

*Lemma 2:* The LM iterates $\lambda_n^i$ converge to optimal value $\lambda^{i,*}$.

*Proof:* The proof is similar to that in [17], Lemma 4 and Corollary 1. Note that the $\tilde{V}_n^i$ and $\lambda_n^i$ iterations are primal-dual iterations. The primal iterations perform relative value iteration and determine a minimum of the Lagrangian with respect to the policy for an almost constant LM. The limiting o.d.e. for the $\lambda_n^i$'s is a steepest ascent for the Lagrangian minimized over the primal variables (See Lemma 4 of [17]), a fact that can be justified by using the 'envelope theorem' [38], which allows one to interchange the 'max' and the gradient operator. By standard results for stochastic gradient ascent for concave functions (see Section 10.2 of [37]), this converges to the optimal LM $\lambda^{i,*}$ as argued in [17] Lemma 4 and Corollary 1. ∎

Lemmas 1 and 2 imply that $(\tilde{V}_n^i, \lambda_n^i) \to (\tilde{V}^i, \lambda^{i,*})$ as required.

The LM iteration can be considered to be a noisy discretization of a limiting o.d.e. which is driven by the shortfall from or the excess over the allowed mean traffic, of the actual mean traffic. Hence it is clear that the convergence of LM to an equilibrium of this o.d.e. implies that the constraints are met with equality.

∎

## B. Queue Stability

In this sub-section we prove (under suitable conditions) that under the strategies learnt by individual users, the queues do remain stable. This is done using a stochastic Lyapunov argument. This does, however, presuppose that the users have learnt the correct strategies already, i.e., the learning scheme has already converged. As noted earlier, this does not prove closed loop stability *with* the learning scheme. Our simulation results, however, indicate a stable behavior even with the learning component kicking in.

Let $\mathcal{V}(\mathbf{q}) = \sum_i q^i$, $\gamma^i = \mathbf{E}[A_n^i]$. Let $\pi$ denote the (unique) stationary distribution of the Markov chain $\{\mathbf{X}_n\}$. Let $R \triangleq \min_{\mathbf{x}} \sum_{\mathbf{x}} \pi(\mathbf{x}) \hat{R}^i(x^i)$.

Recall that $R_n^i$ depends on $(Q_n^i, X_n^i)$. Suppose that $R_n^i = \ell^i(Q_n^i, X_n^i)$ for some $\ell^i(\cdot, \cdot)$. Write $\boldsymbol{\ell}(\mathbf{Q}_n, \mathbf{X}_n)$ for

$[\ell^1(Q_n^1, X_n^1), \ldots, \ell^N(Q_n^N, X_n^N)]$. Now,

$$\mathbf{E}[\mathcal{V}(\mathbf{Q}_{n+1})|\mathbf{Q}_n, \mathbf{X}_n] - \mathcal{V}(\mathbf{Q}_n)$$

$$= \mathbf{E}[\sum_i (Q_{n+1}^i - Q_n^i)|\mathbf{Q}_n, \mathbf{X}_n]$$

$$= -\sum_i U_n^i + \sum_i \gamma^i$$

$$= -\sum_i \ell^i(Q_n^i, X_n^i) I_{\{\ell^i(Q_n^i, X_n^i) \geq \ell^j(Q_n^j, X_n^j), \ i \neq j\}}$$

$$+ \sum_i \gamma^i. \tag{27}$$

If $Q_n^i \uparrow$ then $\ell^i(Q_n^i, X_n^i) \uparrow$ by monotonicity of optimal single user policy in the queue length [10]. It is reasonable to expect that as $q^i \uparrow$,

$$\ell^i(q^i, x^i) \uparrow \hat{R}^i(x^i). \tag{28}$$

This motivates our assumption:

$$\max_i \ell^i(q^i, x^i) > R \tag{29}$$

outside a bounded set of $\mathbf{q}$'s. That is, if queue length(s) blow up, the system at the very least transmits at rate $R$. We also assume:

$$\sum_i \gamma^i < R. \tag{30}$$

This implies that whenever $\mathbf{Q}_n$ is outside a certain large bounded set,

$$\mathbf{E}\left[\mathcal{V}(\mathbf{Q}_{n+1})|\mathbf{Q}_n, \mathbf{X}_n\right] - \mathcal{V}(\mathbf{Q}_n)$$
$$= -R + \sum_i \gamma^i < 0. \tag{31}$$

Thus $\mathcal{V}(\cdot)$ serves as a stochastic Lyapunov function for the queue thus ensuring its stability (i.e., positive recurrence – see [39], Chapter 13). (Note that it suffices to have the Lyapunov function depend on queue lengths alone, as the channel state is in any case finite valued.)

The condition (30) ensures that the minimum of the maximum rate allowed to each user is itself adequate to handle all traffic. Looking at the aggregate queue length process $Q_n := \sum_i Q_n^i$, which satisfies

$$Q_{n+1} = Q_n - \sum_i U_n^i + \sum_i A_{n+1}^i,$$

it is clear that such a condition would be sufficient. Nevertheless it is rather strong and we suspect that it can be significantly weakened.

Under (30), we can say more than just positive recurrence, viz., we characterize the set near which the queue lengths will concentrate, in an approximate way. For technical reasons, for the purposes of this proof we assume that at each time $n$ the base station uses a smooth approximation $\mathbf{F} := [F^1, \ldots, F^N]$ of the maximum function for channel allocation, i.e., it allocates the channel to the user with the highest demand (say, $i$th) with a probability $F^i(\mathbf{R}_n)$ close to one, but does allocate to others with a small but nonzero probability $F^j(\mathbf{R}_n), j \neq i$. Specifically, let $F^i(\mathbf{R}_n) := g(R_n^i)^m / (\sum_j g(R_n^j)^m)$ for some monotone increasing and smooth $g$ and $m >> 1$. Also, we assume that the $\ell^i(\cdot, \cdot)$ above is continuously differentiable. These hypotheses avoid some difficulties in the argument below that would be caused by having to deal with a differential equation with a discontinuous right hand side. This is at the expense of being only approximate.

The queue evolution (2) for user $i$ can be rewritten as:

$$Q_{n+1}^i = Q_n^i + (\gamma^i - F^i(\mathbf{R}_n) R_n^i) + \Big( (A_{n+1}^i - \gamma^i)$$
$$+ (F^i(\mathbf{R}_n) R_n^i - U_n^i) \Big).$$

Let $\widetilde{\mathbf{F}} = [\widetilde{F}^1, \ldots, \widetilde{F}^N]$ be defined by: $\widetilde{F}^i(r) = F^i(r) r^i \ \forall i$. Note that this is continuously differentiable. Then the queue evolution is of the form:

$$Q_{n+1}^i = Q_n^i + (\gamma^i - \widetilde{F}^i(\mathbf{R}_n)) + M_{n+1}, \ n \geq 0, \tag{32}$$

where $\{M_n\}$ is a martingale difference sequence.

Let $\mathbf{q} = [q^1, \ldots, q^N]$. Define $\widehat{\mathbf{F}} = [\widehat{F}^1, \ldots, \widehat{F}^N]$ by:

$$\widehat{F}^i(\mathbf{q}) = \sum_{\mathbf{x}} F^i(\boldsymbol{\ell}(\mathbf{q}, \mathbf{x})) \ell^i(q^i, x^i) \pi(\mathbf{x}).$$

Consider the scaled version of (32), given by

$$Q_{n+1}^i = Q_n^i + \eta[(\gamma^i - \widetilde{F}^i(\mathbf{R}_n)) + M_{n+1}], \ n \geq 0, \tag{33}$$

for a small $\eta > 0$. If we consider smaller and smaller time slots of width $\eta$ with $\gamma^i$, and $\widetilde{F}^i$ being 'rates per unit time' rather than 'per slot' quantities, then we obtain the 'fluid' approximation of (32), which is given by the o.d.e.

$$\dot{q}^i(t) = \gamma^i - \widehat{F}^i(\mathbf{q}(t)), \tag{34}$$

restricted to the positive orthant (i.e., it follows the projected dynamics on the boundary of this orthant).

*Lemma 3:* The trajectories of the o.d.e. in (34) for each $i$ approach a bounded set.

*Proof:* The proof is similar to the stochastic Lyapunov argument above (Refer to (31)). Let $\Omega_n \triangleq \{$ maximizers of $i \to R_n^i\} \implies$ for $m >> 1, F^i(\mathbf{Q}_n, \mathbf{X}_n) \approx \frac{1}{|\Omega_n|} I_{\{i \in \Omega_n\}}$. Let $\mathcal{V}(\mathbf{q}) \triangleq \sum_i q^i$. Thus

$$\dot{\mathcal{V}}(\mathbf{q}(t)) = \sum_i \gamma^i - \sum_i \widehat{F}^i(\mathbf{q}(t))$$

$$\approx \sum_i \gamma^i - \sum_x \pi(\mathbf{x}) \ell^{i^*(t,x)}(q^{i^*(t,x)}(t), x^{i^*(t,x)}),$$

where $i^*(t, x) \triangleq$ the maximizer of $i \to \ell^i(q^i(t), x^i)$, i.e., $i^*(t, x)$ is index of the user that bids the highest rate. If any $q^i(t)$ becomes very high, then one expects the users to transmit at highest possible rate in all channel states, and the system at the very least transmits at rate $R$. Hence, r.h.s. $\leq \sum_i \gamma^i - R$, thus leading to

$$\dot{\mathcal{V}}(\mathbf{q}(t)) \leq \sum_i \gamma^i - R < 0.$$

Thus $\mathcal{V}(\cdot)$ serves as a Lyapunov function for the o.d.e., leading to bounded trajectories. ■

Recall that an o.d.e. $\dot{x}(t) = f(x(t))$ is cooperative if $\frac{\partial f_i}{\partial x_j} > 0$ for $i \neq j$ (more generally, $\geq 0$ and the Jacobian matrix with diagonal elements replaced by zero is irreducible) [40].

*Lemma 4:* $\frac{\partial \widehat{F}^i}{\partial q^j} < 0$ for $i \neq j$, i.e., (34) is a cooperative dynamics.

*Proof:* From [7], [10], we know that the optimal policy is increasing in queue length for the single user case with i.i.d. channel fading and arrival processes. Thus, increase in the queue length $Q^j$ for a user $j$ results in an increase in $R^j$ as computed by user $j$. As per our user selection policy at the base station, this results in a decrease in probability of channel allocation $F^i(\cdot), i \neq j$, and hence $\widehat{F}^i(\cdot)$ for user $i$. ∎

We thus have the following lemma.

*Lemma 5:* The trajectories of the o.d.e. in (34) converge to an equilibrium for almost all initial conditions.

*Proof:* Boundedness of the iterates as proved in Lemma 3 and the cooperative property as proved in Lemma 4 allows us to apply Hirsch theorem (Theorem 4.1, p. 435, of [40]). It follows that the trajectories converge for almost all initial conditions. ∎

*Lemma 6:* With probability $1 - O(\eta)$, $\{\mathbf{Q}_n\}$ in (32) will concentrate near the set of equilibria $B$ of (34).

*Proof:* As seen above, $\mathcal{V}(\cdot)$ serves as a *stochastic* Lyapunov function for the Markov chain $\{\mathbf{Q}_n\}$, implying its positive recurrence. In particular, it will be asymptotically stationary. Thus it suffices to consider the stationary chain. Consider two time slices at $n$ and $n + M$ resp., $n, M > 0$, of the stationary chain. The distribution $\mu_s$ of the queue lengths is the same at both times by stationarity. Let $\eta > 0$ as above and pick $K \subset (\mathcal{Z}^+)^N$ such that $\mu_s(K) > 1 - \eta$. Let $B^\epsilon$ denote the $\epsilon$-neighborhood of $B$. Then $\mu_s(\cup_i \{q : q(0) = q, q(t) \in B^\epsilon \ \forall t \geq i\}) = 1$ by Lemma 5. Thus we can pick an $M \geq 1$ such that with probability $\mu_s(\{q : q(0) = q, q(t) \in B^\epsilon \ \forall t \geq M\}) > 1 - \eta$. Let $D \triangleq \{q : q(0) = q, q(t) \in B^\epsilon \ \forall t \geq M\}$. Now consider (33) as a constant stepsize stochastic approximation iteration for small $\eta$. The convergence analysis for constant stepsize stochastic approximation algorithms of Chapter 9, [37], implies that

$$E[d(\mathbf{Q}_{n+\lceil \frac{M}{\eta} \rceil}, B^\epsilon)^2 | \mathbf{Q}_n \in D] \leq C\eta \qquad (35)$$

for some $C > 0$, $d(\cdot, B^\epsilon)$ being the Euclidean distance from $B^\epsilon$. It follows that under the stationary law,

$$\begin{aligned} &P(d(\mathbf{Q}_{n+\lceil \frac{M}{\eta} \rceil}, B^\epsilon) \geq \epsilon) \\ &= P(d(\mathbf{Q}_{n+\lceil \frac{M}{\eta} \rceil}, B) \geq 2\epsilon) \\ &\leq \eta + O\left(\frac{\eta}{\epsilon^2}\right), \end{aligned}$$

where we use stationarity, (35) and the Chebyshev inequality to get the above inequality. ∎

The equilibria $q^*$ of the limiting o.d.e. correspond to situations where

$$\gamma^i = \widehat{F}^i(\mathbf{q}^*) \ \forall i, \qquad (36)$$

i.e., the mean arrivals and departures balance out, as they should. The discussion above then shows that the stochastic behavior fluctuates around this. Solution of (36) for $m >> 1$ thus allows us to make predictions about the queue behavior.

### C. Some Optimality Properties

We now prove that the base station user selection algorithm minimizes the long term sum of maximum rates. Moreover, we argue that the average power vector is a point on the Pareto frontier and hence it is Pareto optimal. Furthermore, we also argue that the scheme is 'fair' in the sense that users requiring larger system resources have to pay a higher 'price', i.e., they expend more power.

*1) User Selection Algorithm:* The base station faces a 'restless bandit' problem [41] with partial observations, since the base station has a knowledge of $\mathbf{X}_n$ and not of $\mathbf{Q}_n$. This problem is hard [42]. In the user selection policy suggested by us, the base station employs a greedy index policy treating the observed bids as indices. Here, we prove the optimality of the user selection policy at the base station for a certain cost criterion, *given* the user behavior.

*Lemma 7:* The base station user selection policy minimizes:

$$\limsup_{M \to \infty} \frac{1}{M} \sum_{n=1}^{M} \max_i \ell^i(Q_n^i, X_n^i). \qquad (37)$$

*Proof:* Let $\Pi(\mathbf{Q}, \mathbf{X})$ be stationary distribution of the joint queue and channel state under the user selection policy proposed in the paper. Taking stationary expectations on both sides of the recursion

$$\sum_i Q_{n+1}^i = \sum_i Q_n^i - \sum_i U_n^i + \sum_i A_{n+1}^i$$

allows us to write the following:

$$\begin{aligned} \sum_i \gamma^i &= \sum_{\mathbf{q}, \mathbf{x}} \Pi(\mathbf{q}, \mathbf{x}) \sum_i I_{\{\ell^i(q^i, x^i) > \ell^j(q^j, x^j)\}} \ell^i(q^i, x^i) \\ &= \sum_{\mathbf{q}, \mathbf{x}} \Pi(\mathbf{q}, \mathbf{x}) \max_i (\ell^i(q^i, x^i)). \qquad (38) \end{aligned}$$

Let $\tilde{\Pi}(\mathbf{q}, \mathbf{x})$ be the stationary distribution of joint queue and channel state induced by some other stabilizing policy that does not necessarily schedule the user bidding the highest rate in a slot. (It is assumed that the other policy is a stabilizing policy, but then it is clear that unstable policies are not contenders for optimality anyway.) Again, we have:

$$\sum_i \gamma^i = \sum_{\mathbf{q}, \mathbf{x}} \tilde{\Pi}(\mathbf{q}, \mathbf{x}) \sum_i \chi^i(\mathbf{q}, \mathbf{x}) (\ell^i(q^i, x^i)), \qquad (39)$$

where $\chi^i(\mathbf{q}, \mathbf{x})$ is the probability of scheduling user $i$ in state $(\mathbf{q}, \mathbf{x})$ under the other stabilizing policy; $\chi^i(\mathbf{q}, \mathbf{x}) \geq 0$, $\sum_i \chi^i(\mathbf{q}, \mathbf{x}) = 1$. This implies that:

$$\begin{aligned} &\sum_{\mathbf{q}, \mathbf{x}} \Pi(\mathbf{q}, \mathbf{x}) \left( \max_i (\ell^i(q^i, x^i)) \right) \\ &= \sum_{\mathbf{q}, \mathbf{x}} \tilde{\Pi}(\mathbf{q}, \mathbf{x}) \left( \sum_i \chi^i(\mathbf{q}, \mathbf{x}) (\ell^i(q^i, x^i)) \right) \\ &\leq \sum_{\mathbf{q}, \mathbf{x}} \tilde{\Pi}(\mathbf{q}, \mathbf{x}) \left( \max_i (\ell^i(q^i, x^i)) \right). \qquad (40) \end{aligned}$$

Hence the policy minimizes (37) over all stable stationary policies. From standard results from Markov decision theory ([30], Chapter 8), this implies the claim. ∎

*Remark 3:* Note that $\ell^i(\cdot, \cdot)$ is the user $i$'s policy, i.e., the function that maps the state of user $i$ to the rate it bids, which in turn determines its power expenditure. Thus the minimization objective of the base station, which seeks to minimize the average *maximum* rate requested per slot, aids in conserving user power. This is not enough to show that power consumption *per user* is minimized. However, we argue that the scheme is Pareto optimal.

*2) Pareto Optimality of Power Vector:* Once the users begin transmitting at stable transmission power schedule, in order to reduce the power consumption of say, user $i$, the base station has to reduce the average rate with which user $i$ transmits. Since the delay constraint of user $i$ must be satisfied, this can be done by increasing the fraction of slots allocated to that user. This results in decreasing the fraction of slots allocated to some other user(s), say, user $j$. Now, if user $j$ has to satisfy its delay constraint, it has to increase the rate at which it transmits, thus increasing its power expenditure. Thus, once the system has stabilized, reduction in the power expenditure of one user is possible only at an expense of increase in the power expenditure of some other user(s). Therefore the long term power expenditure vector is a point on the Pareto frontier, i.e., the scheme is Pareto optimal.

### D. Game Theoretic Interpretation

Here, we provide a game theoretic interpretation of the scheme and argue that it is incentive compatible.

*1) Multi-user Penalty:* Consider a user $i$. Fix a Markov policy learnt by users $j \neq i$ and the base station policy. As per our scheme, user $i$ solves an average cost MDP with running cost function $P(x, r) + \lambda^i(q - \bar{\delta}) = P(x, r) + \lambda^i_*(\frac{\lambda^i}{\lambda^i_*}(q - \bar{\delta}))$, where $\lambda^i_*$ is the correct LM for the constrained MDP with the objective of minimizing $\limsup_{M \to \infty} \frac{1}{M} \sum_{n=1}^{M} P(Q_n^i, R_n^i)$ subject to $\limsup_{M \to \infty} \frac{1}{M} \sum_{n=1}^{M} Q_n^i \leq \bar{\delta}^i$. This can be interpreted as there being an appropriate 'scaling' of the LM because of the presence of multiple users. This is the penalty of operating in a multi-user environment. The upward scaling of the LM would naturally result in a higher power consumption as compared to the single user facing the aforementioned constrained MDP. We call the ratio $\frac{\lambda^i}{\lambda^i_*}$ the *multi-user penalty*. It is clear that each user is optimal for this modified MDP when the policies of other users and the base station are fixed. On the other hand, we have already seen that the base station uses a policy that is optimal for cost (37) when the user policies are fixed. Thus this is a Nash equilibrium for a certain stochastic dynamical game. In fact, it is an instance of the economists' notion of a *Markov equilibrium* (See, e.g., Chapter 13 of [43]).

*2) Incentive Compatibility of the Scheme:* The users always attempt to keep their average queue lengths close to their respective queue length constraints, i.e., the constraint is satisfied with equality. This is because the user's objective is to minimize the power expenditure. By asking for a higher rate than what is required, a user might achieve an average queue length that is much lower than the queue length constraint. However, as proved in [7] for single user policy, since the power is an increasing convex function of the rate, the rate should be kept as low as possible so that the average queue length is as large as possible (in this case, equal to the queue length constraint) in order to save power. Thus users will transmit at a rate such that the average throughput achieved by them is just sufficient to meet the delay constraint with equality, implying that there is no incentive for the users to lie and ask for an unnecessarily high rate. This establishes the incentive compatibility of the scheme.

*3) Fairness:* The scheme is 'fair' in the sense that users' power expenditure is commensurate with their fraction of time slots requirement. A certain user having a high arrival rate or stringent delay constraint or poorer channel condition requires higher fraction of slots in order to satisfy the delay requirement. Such a user must consistently bid higher rates in order to obtain a higher fraction of time slots and thus ends up 'paying' more in terms of its long term power expenditure.

## VI. EXPERIMENTAL EVALUATION

We demonstrate the performance of our algorithm under the IEEE 802.16 [1] framework through simulations in a discrete event simulator. Specifically, we intend to demonstrate the following:

1) The algorithm satisfies the users' delay constraints.
2) The algorithm is efficient in terms of the power consumed for each user. This is demonstrated by comparing the power consumed under our algorithm with that under M-LWDF scheduler [24].
3) Performance of the algorithm under different information accuracies.

### A. Simulation Setup

We consider uplink (UL) transmissions in the *residential* scenario as in [44]. Internet traffic is modeled as a web traffic source [44], [45]. Variable sized packets are generated at the application layer. Packet sizes are drawn from a truncated Pareto distribution. This distribution is characterized by three parameters: shape factor $\xi$, mode $\upsilon$ and cutoff threshold $g$. The probability that a packet can have a size $y$ can be expressed as:

$$f_{TP}(y) = \frac{\xi \cdot \upsilon^\xi}{y^{\xi+1}}, \quad \upsilon \leq y < g$$
$$f_{TP}(y) = \nu, \quad y \geq g, \quad (41)$$

where $\nu$ can be calculated to be equal to:

$$\nu = (\frac{\upsilon}{g})^\xi, \; \xi > 1. \quad (42)$$

We choose shape factor $\xi = 1.2$, mode $\upsilon = 2000$ bits, cutoff threshold $g = 10000$ bits, which provides us with an average packet size of 3860 bits. In each time frame, we generate the arrivals for all the users using Poisson distribution. Arrivals are generated in an i.i.d. manner across frames. We divide the packets into fragments at the MAC layer with each fragment being of size $\tau = 2000$ bits. Fragments of size less than 2000 bits are padded with extra bits. Since all fragments are of equal size, we determine the transmission rate for users in terms of number of fragments. We simulate a Rayleigh fading channel[3] for each user. For a Rayleigh model, channel state $X^i$ is an exponentially distributed random variable with mean $\alpha^i$ and probability density function expressed as $f_X(x) = \frac{1}{\alpha^2} \exp\left(\frac{-x^2}{2\alpha^2}\right)$, $x \geq 0$. We assume that the power required for transmitting $z$ fragments of size $\tau$ bits when the channel state is $x$ is $P(x, z\tau) = \frac{WN_0}{x}\left(2^{\frac{z\tau}{W}} - 1\right)$ where $W$ is the bandwidth and $N_0$ is the power spectral density of the additive white Gaussian noise at the receiver. We assume that the product $WN_0$ is normalized to 1. We measure the sum of queuing and transmission delays of the packets and ignore the propagation delays. In all the scenarios described below, a single simulation run consists of running the algorithm for 100000 frames and the results are obtained after averaging over 20 simulation runs. We discretize the channel into eight equal probability bins, with the boundaries specified by $\{$ $(-\infty, -8.47$ dB$)$, $[-8.47$ dB$, -5.41$ dB$)$, $[-5.41$ dB$, -3.28$ dB$)$, $[-3.28$ dB$, -1.59$ dB$)$, $[-1.59$ dB$, -0.08$ dB$)$, $[-0.08$ dB$, 1.42$ dB$)$, $[1.42$ dB$, 3.18$ dB$)$, $[3.18$ dB$, \infty$ $)$ $\}$. For each bin, we associate a channel state and the state space $\mathcal{X} = \{$ $-13$ dB, $-8.47$ dB, $-5.41$ dB, $-3.28$ dB, $-1.59$ dB, $-0.08$ dB, $1.42$ dB, $3.18$ dB$\}$. This discretization of the state space of $X^i$ has been justified in [8]. We assume $N = 20$, i.e., a system with 20 users and thereby 20 UL connections. We assume that the number of users does not change during the course of simulations. In the symmetric case, all 20 users have same parameters while in the asymmetric case, users are divided into two groups (Group 1 and Group 2) of 10 users each with different parameters. We have simulated both symmetric and asymmetric cases. However, due to space constraints, we include the results for the asymmetric case only.

### B. Satisfaction of Delay Constraint

In this sub-section, we demonstrate that the delay constraint of each user is satisfied for various constraints and average channel conditions.

*Scenario 1:* In this scenario, we demonstrate that the algorithm satisfies the various user specified delay constraints. In each frame, arrivals are generated with a Poisson distribution with mean 0.1 packets/msec. Packet lengths are Pareto distributed with parameters detailed above. This results in an arrival rate of 0.386 Mbits/sec/user. We choose $\alpha^i = 0.4698(-3.28$ dB$)$ $\forall i$. In each slot, we generate $X^i$ using exponential distribution with mean $\alpha^i$. We determine the channel state based on the bin that contains $X^i$ as explained above.

We perform multiple experiments. The delay constraints of the users in Group 1 are fixed at 100 msec in each experiment, while the delay constraints of the users in Group 2 are fixed at 25, 50, 75, 100, 125, 150, 175 msec in successive experiments. It can be observed from Figure 3 that the delay constraints are satisfied in for all the constraints. Moreover, from Figure 4, it can be observed that the power expended is a convex decreasing function of the delay constraint imposed by the user. Larger delay constraint implies that lesser power is required to satisfy the constraint.

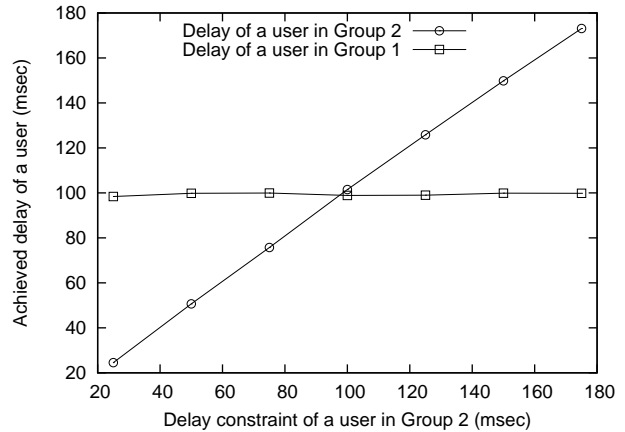

Fig. 3. Achieved delay of a user with specified delay constraints

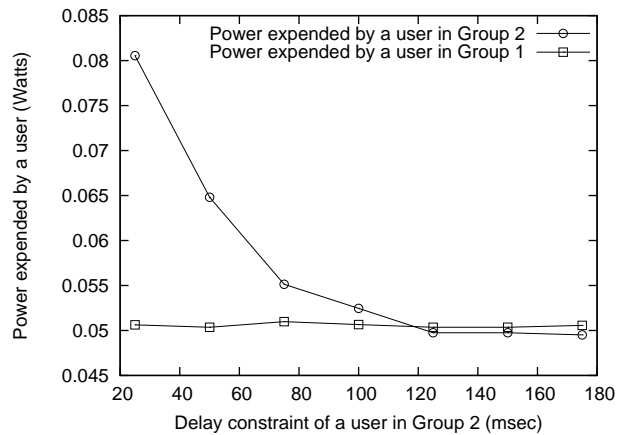

Fig. 4. Power expended with specified delay constraints

*Scenario 2:* In this scenario, we demonstrate that the algorithm satisfies the user specified delay constraints for various channel conditions. We consider the asymmetric case where we maintain the average channel state for users in Group 1 constant for all the experiments, i.e., $\alpha^i = -3.28$ dB, $i \in 1, \ldots, 10$. For the users in Group 2, $\alpha^i$ for $i \in 11, \ldots, 20$, the average channel state is fixed at $\alpha^i = -13$ dB, $-8.47$ dB, $-5.41$ dB, $-3.28$ dB, $-1.59$ dB, $-0.08$ dB, $1.42$ dB, in successive experiments. Average delay suffered by a user in Group 1 and in Group 2 and power consumed by them for the two cases are plotted in Figures 5, and 6 respectively. From Figure 5, it can be observed that the scheme is able to satisfy the delay constraints above a certain average channel state. The maximum power with which the users can transmit in any slot determines the capacity of the system. If the maximum power is high, the scheme is able to satisfy the delay constraints even for poor channel states. Thus, the maximum power determines the average channel state above which the delay constraints are satisfied. From Figure 6 it can be observed that better channel conditions result in much lesser power being required for satisfying the delay constraints.

### C. Comparison with M-LWDF

Here, we compare the power consumed by our algorithm with the M-LWDF scheme [24]. The arrival rate of all the users is fixed at 0.2702 Mbits/sec (0.07 packets/msec) for all the experiments while $\alpha^i = 0.4698(-3.28$ dB$)$ $\forall i$. We first determine the average delay experienced by the users with different maximum power constraints (1.5 - 4.5 Watts) under M-LWDF scheme. We keep track of the power expended by each user under the M-LWDF scheme. The delays experienced under M-LWDF scheme are set as delay constraints for the proposed scheme. Figure 7 demonstrates that the proposed scheme satisfies the delay constraints. The comparison for power consumed under the two schemes is plotted in Figure 8. It can be seen from this figure that the proposed scheme consumes much less average power than the M-LWDF scheme.

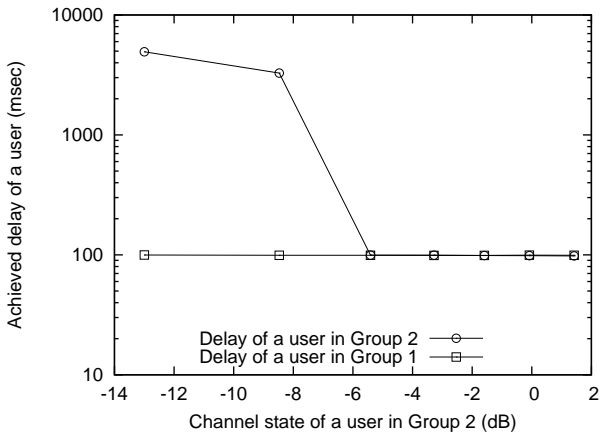

Fig. 5. Achieved delay of a user with varying channel conditions

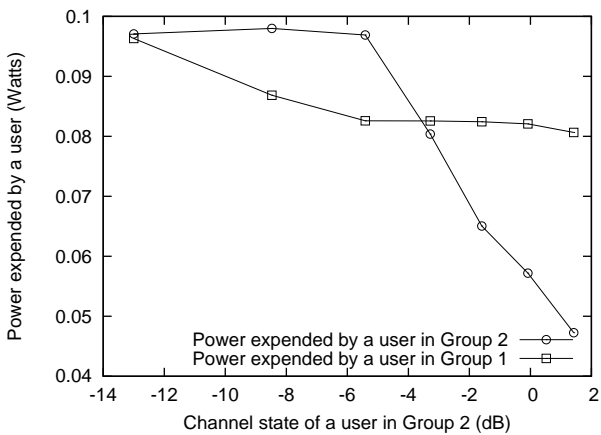

Fig. 6. Power expended with varying channel conditions

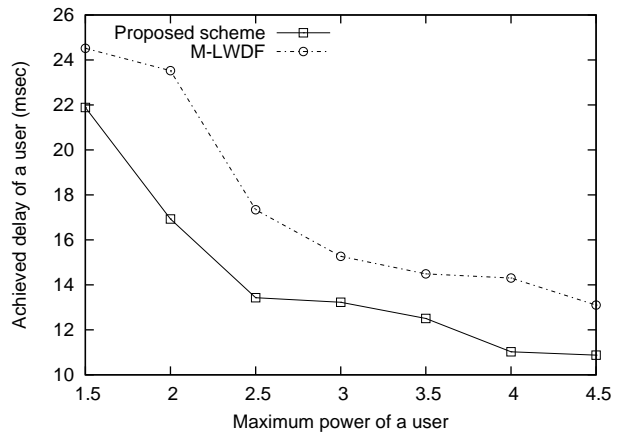

Fig. 7. Comparison with M-LWDF Scheduler - delay

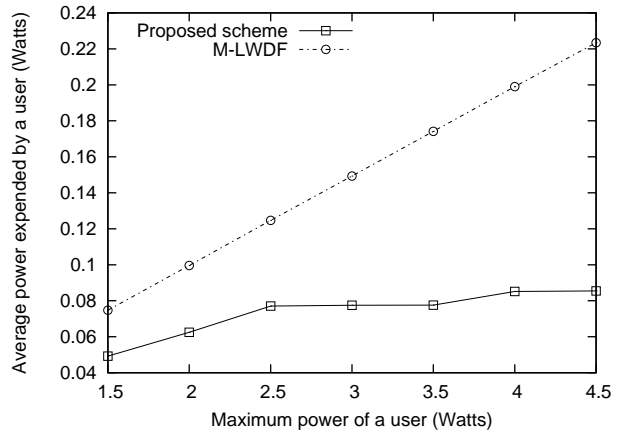

Fig. 8. Comparison with M-LWDF Scheduler - power

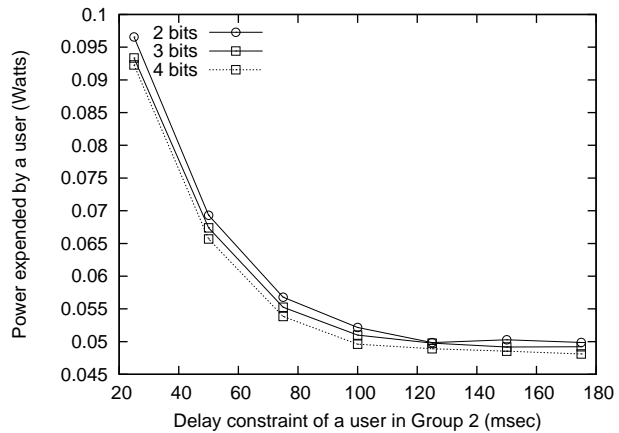

Fig. 9. Power consumed for various information accuracies

### D. Performance under Different Information Accuracies

Here, we study the performance of our scheme under different accuracies of information. Specifically, we determine the power consumption when 2 bits, 3 bits and 4 bits are employed in order to convey the rate information from the users to the base station. The parameters for the users are same as those in Scenario 1 of Section VI-B. The results are plotted in Figure 9. It can be seen that as the information accuracy is increased, the power consumption reduces.

*Remark 4:* It is apparent that there exists a tradeoff between the information accuracy and the resources (power, bandwidth)

consumed in conveying the information. We are currently investigating this tradeoff.

## VII. CONCLUSIONS

In this paper, we have proposed a novel scheduling algorithm for minimizing the average power of each user subject to individual delay constraint in a multi-user uplink system. The primary difficulty in numerically determining an optimal policy for this problem is the large state space. To address this difficulty, we have proposed a novel extension of single user optimal algorithm to the multi-user setting. In our approach, the users can be thought of as bidding their rates to the base station which then schedules the user bidding the highest rate. We note that it is not in the interest of users to bid unnecessarily higher rates as that might result in higher power consumption. We prove analytically that the proposed algorithm ensures user queue stability if the learning scheme converges and vice versa, and if so, satisfies the delay constraints of the users. Another advantage of our approach is that it does not require an explicit knowledge of the probability distributions of channel state and arrival processes. The algorithm is computationally efficient and has low communication overhead. It thus provides a powerful framework for uplink scheduling. Interesting future directions are to explore a network situation, and on a different note, to provide a more complete theoretical analysis.